\documentclass[12pt,preprint]{aastex}

\shorttitle{Evolution of coated grains in spiral shocks}
\shortauthors{Podolak, Mayer, and Quinn}

\usepackage{graphicx}

\begin{document}

\title{Evolution of Coated Grains in Spiral Shocks of  Self-Gravitating Protoplanetary Disks}

\author{M. Podolak$^1$, L. Mayer$^2$, and T. Quinn$^3$}\affil{$^1$Dept. of Geophysics \& Planetary Science, Tel Aviv University, Tel Aviv Israel 69978\\$^2$Institute for Theoretical Physics, University of Zurich, Winterthurerstrasse 190, 8057 Zurich, Switzerland\\$^3$Dept. of Astronomy, University of Washington, Seattle, WA 98195}

\begin{abstract}

We investigate the evolution of grains composed of an ice shell surrounding an olivine core as they pass through a spiral shock in a protoplanetary disk.  We use published three-dimensional radiation-hydrodynamics simulations of massive self-gravitating protoplanetary disks to extract the thermodynamics of spiral shocks in the region between $10$ and $20$ AU from the central star. As the density wave passes, it heats the grains, causing them to lose their ice shell and resulting in a lowering of the grain opacity.  In addition, since grains of different sizes will have slightly different temperatures, there will be a migration of ice from the hotter grains to the cooler ones.  The rate of migration depends on the temperature of the background gas, so a grain distribution that is effectively stable for low temperatures, can undergo an irreversible change in opacity if the gas is temporarily heated to above $\sim 150$\,K.  We find that the opacity can drop more, and at a significantly faster
rate throughout the spiral shocks relative to the prediction of standard dust grains models adopted in hydrodynamical calculations of protoplanetary disks. 
This would lead to faster gas cooling  within spiral arms. We discuss the implications of our results on the susceptibility of disk fragmentation into sub-stellar objects at distances of a few tens of astronomical units.

\end{abstract}


\section{Introduction}

Protostellar disks likely undergo one or more phases of gravitational instability during the first $10^5$ to $10^6$ years of their life \citep{vorobyov10b, vorobyov10a, hayfield10}. Two factors determining the susceptibility to gravitational instability, as well as its ultimate outcome, are the mass loading from the molecular cloud envelope \citep{boley09,rafikov09,boleyetal10} and how fast the gas is able to cool and dissipate the energy generated by compression and shocks in the spiral arms \citep{riceetal05,boleyetal07,mayer07,cossins10,merubate10}, either by radiative cooling alone, or by a combination of radiative cooling and convective cooling \citep{boss03, durisetal07, mayer10}. Whether gravitational instability results in a self-regulated, marginally unstable state with long-lived spiral structure, or whether self-regulation is broken and fragmentation takes place within  spiral arms, producing sub-stellar objects on brown dwarf or giant planet scales \citep{stamatellos09, boleyetal10,mayer10}, depends on the ability to attain a gas cooling timescale comparable to, or shorter than, the local orbital timescale. Both radiative and convective cooling  ultimately depend on the opacity in the gas. Therefore it is of pivotal importance to investigate, in detail, how the opacity might change within spiral shocks.

At the relatively low temperatures of protoplanetary disks reached at ten or more AU from the star the major source of the opacity is dust grains, and different prescriptions are used to calculate their contribution. Some of these prescriptions are normally used in the 3-D hydrodynamical simulations of gravitationally unstable protostellar disks \citep{d'alessio01} but it is unclear whether they account for all the important effects on the grains as they undergo rapid temperature and density variations within the spiral shocks.  

In this work we explore, in detail, the effect of a varying background temperature and gas density on the grain size distribution and its resultant opacity.  In particular we consider the time-dependent density and temperature evolution in strong spiral shocks appearing in gravitationally unstable protoplanetary disks that attain a minimum Toomre $Q$ parameter just above the value needed for fragmentation [the necessary condition for fragmentation is that the minimum $Q$ be $1.4$ or lower - see e.g \citep{durisetal07}].  We extract such information from one of the 3-D SPH simulations presented in \cite{mayer07}, which included radiative transfer with the flux-limited diffusion approximation. 

The paper is organized as follows: First we describe the model adopted for the three-dimensional spiral shocks and for dust grains, then we describe our results on the evolution of dust grain opacity in our reference dust model, analyzing different locations throughout the spiral shock. We show results for the disk midplane first, and then for regions well above the midplane, including the effect of irradiation from the star. Furthermore, we investigate the effect of changing parameters in our model for dust, such as the dust-to-gas ratio in  the spiral shocks, the water content of grains, and the distribution of grain sizes.  Subsequently, we explore a model in which dust grains are composite. We conclude the paper with a discussion of the implications of our results for disk cooling and fragmentation into sub-stellar objects.

\section{Model setup}

The thermodynamics of spiral shocks adopted throughout the paper is extracted from the simulation shown in the bottom left corner of Fig.\,1 of \cite{mayer07}, that led to long-lasting strong spiral structure (the minimum Toomre $Q$ parameter was hovering around $\sim 1.5$ for many orbital times) but not to fragmentation [see \cite{mayer07} for details on the parameters of this particular run].  Fragmentation did not occur because the gas was unable to radiate on a sufficiently short timescale.  

The profiles are computed in thick annuli centered on the density maximum of one of the spiral shocks (which also roughly corresponds to the temperature maximum). In the figure, the circle marks the radial distance at which all annuli are centered.  However, in order to avoid diluting the effect of the spiral arm, the profiles are not computed by azimuthally averaging over a $2\pi$ annulus (i.e. over the whole disk), rather they are azimuthally averaged across a 3-D rectangular box with an initial height of $0.75$\,AU and an initial azimuthal extent of 4\,AU, as shown in grey. At any time the box is chosen in such a way as to encompass the same particles within the chosen section of the spiral shock, exploiting the Lagrangian nature of SPH. 
Therefore the box changes in size as time progresses because particles diffuse in space as the shock widens (see Figure 1). 
The different annuli have increasing radial width up to edges of the box, so that they probe conditions increasingly further away from the density maximum, and, in order to simplify the comparison the width of each annulus is kept constant while 
the box changes in size (we verified that we are able to follow most of the particles initially identified within the box).

We consider on purpose a vigorously unstable but non-fragmenting disk case because it should represent a case in which a sudden variation in the cooling timescale, eventually triggered by opacity changes, might rapidly bring the disk from a self-regulated state to rapid fragmentation \citep[see][]{merubate10, cossins10}.  At the location of the spiral shock the disk reaches the minimum Toomre $Q \sim 1.5$ (about $12$\,AU from the center).  In the simulation the disk was growing steadily in mass over hundreds of orbital times; a crude attempt to model accretion from the surrounding molecular cloud core. At the chosen stage of disk evolution the disk has a total mass of about $\sim 0.12 M_{\odot}$ out to $30$\,AU, i.e slightly above $10\,\%$ of the mass of the central star (which has a mass of $1\,M_{\odot}$).  

The SPH simulations of \cite{mayer07} used the opacity model of \cite{d'alessio01}, which was, in turn, based on the model of \cite{pollack94}. \cite{pollack94} considered both spherical grains and spheroids where the axial ratio was 4:1. They also considered details of the chemical composition of the different components.  They investigated the case where each component is segregated in a separate class of grains, as well as the case where the grains were composites of several species, produced by the coagulation of smaller grains.  A possibility that was not considered, however, is the case where the grains are composed of a refractory core surrounded by an ice shell.  Such a situation might occur if the grains were in a spiral shock, where the heating would evaporate the water ice.  

As the gas cools the water vapor would recondense on the silicate nuclei to form an ice mantle.  In such a case, we might expect that the change of opacity with temperature would be different from the changes found by \cite{d'alessio01} and \cite{pollack94}. This could have important consequences for the stability of the disk models.  We investigate this case in the present paper.   

To model this process, we consider a distribution of silicate cores.  We assume that the number of cores per unit volume with radii between $a$ and $a+da$ is given by a power law: 
\begin{equation}\label{eq:sizedist}
n(a)da=a^{-\xi}da
\end{equation}
where $\xi$ is a constant.  The standard case we examine will have a ratio of silicates to hydrogen/helium of $4.23\times 10^{-3}$ by mass.  We also assume, following \cite{d'alessio01}, that there is water present and that the mass ratio of water to silicates is 2.13.  

We use a temperature-density relationship in the gas which is relevant for spiral shocks in self-gravitating protoplanetary disks, derived for the case shown in Fig.\,\ref{shockfig1}, and follow the evolution of the grain opacity.  Unless otherwise specified, the water is assumed to be in vapor form initially, even though, for most of the cases considered, this vapor will be supersaturated by many orders of magnitude.  We consider only inhomogeneous nucleation, that is, that the vapor will only condense onto the silicate cores.  Once the silicate cores are introduced, the condensation of water vapor is very fast.  Except for a very short time near the beginning of the simulation, the difference between the case where all the water is in the form of vapor initially and the case where all the water is in the form of an ice shell around the silicate core is negligible.

Marginally unstable protoplanetary disks, which do not undergo fragmentation but rather self-regulate to a minimum Toomre $Q\sim 1.5-1.6$, exhibit a nearly stationary spiral pattern \citep{boleydur08}. In the \cite{mayer07} simulations the dominant spiral mode was $m=3$ over several tens of orbital times. Spiral structure is indeed long-lived in a marginally unstable, self-regulated state.

Locally, the gas density and temperature undergo maxima and minima over time as trailing density waves propagate through the disk, with a timescale of order the local orbital time, and the gas shocks in correspondence to the density jumps.
For our analysis we choose to describe a time window also of order the orbital time at the radius where shocks normally reach their maximum strength, which is of order $12-15$\,AU (at such distances the orbital time is of $30-45$\,yr).  While we have chosen a particular phase of the disk evolution to compute the shock profiles, this choice is by no means special because, as we just mentioned, the spiral pattern is nearly stationary in disks that are in a self-regulated state.

We utilize the structural data for one of the three spiral shocks in particular
(see Fig.\,\ref{shockfig1}) after having verified that the relevant conditions, namely temperature and density evolution, are very similar among the three. We note that disks with a more complicated, higher order, spiral pattern might exhibit more differences between shocks, as these might appear over a wider range of distances in the disk where both background density and temperature as well as shear can vary substantially [compare for example the case of lighter, colder disks considered in \cite{mayer04}].

The temperature-density profiles of the shock as a function of time are shown in Fig.~\ref{shock} for three positions through the spiral arm at the disk midplane; distances are computed with respect to the density maximum along the spiral arm (we average between the conditions at either side of the spiral arm for simplicity since no remarkable temperature or density differences are seen).

For each of these cases we have computed the evolution of the temperature of a grain as a function of time.  At any instant the grain is heated by absorbing ambient radiation from the surroundings, and by latent heat released from the condensation of water vapor onto the grain.  It is cooled by re-radiation and by water ice sublimation.  Contact with the gas is also included, and acts to either cool or heat the grain depending on the sign of the temperature difference between them.  Details of the energy balance calculation are given in \cite{mekler94}, \cite{podmek97}, and \cite{podolak04}.  In computing the cross section for absorption and radiation, we used the FORTRAN code DMILAY.F developed by W. Wiscombe of the Goddard Space Flight Center \footnote{DMILAY.F can be downloaded from ftp://climate1.gsfc.nasa.gov/wiscombe/Single\_Scatt/Coated\_Sphere/}.  It is based on the theory presented in \cite{toonack81}.  This code computes the scattering and absorption cross sections for a spherical grain composed of a core and a shell.

The grain size distribution in our baseline case has $\xi=3.5$, a minimum radius, $a_{min}=10^{-6}$~cm, and a maximum radius, $a_{max}=0.1$~cm, as in \cite{d'alessio01}.  This distribution can be well represented with 5 size bins.  The bins are chosen to divide the logarithm of the size range equally.  The grain size associated with each bin is that of the (logarithmic) center of the bin, and the number of grains in the bin are taken to be the total mass of grains in the size range represented by the bin divided by the mass of the representative grain for that bin.  In our case the silicate core sizes are: $2.21\times 10^{-6}$, $2.21\times 10^{-5}$, $2.21\times 10^{-4}$, $2.21\times 10^{-3}$, and $2.21\times 10^{-2}$\,cm, respectively.

\section{Results I - Opacity Change in the Standard Case}

When the silicate cores are initially inserted into the gas, there is a rapid condensation of the vapor onto the grains.  The background gas is highly supersaturated, so the water condensation flux onto the grains is much higher than the water sublimation flux off the grains.  The net growth rate of the grain is nearly independent of the grain temperature, and of the grain radius.  As a result all the grains acquire an ice shell of the same thickness.  The behavior of the grain opacity (Rosseland mean) as a function of time at different parts of the density wave is shown in Fig.\,\ref{opact}.

The ice shell forms within minutes and the increased radius of the grains causes the grain opacity to rise from around 0.5 to around 3~cm$^2$~g$^{-1}$.  The saturation of the water vapor in the gas phase drops quickly to 1.  At this point the grains are all nearly at the same temperature as the background gas, although a slight difference in temperature between the different grain sizes does exist.  Typically the smallest grains are very near the temperature of the gas, while the larger grains are a fraction of a degree cooler than the gas.  The difference is usually of the order of 10\,mK although it can be as much as 0.3\,K when rapid condensation of water vapor onto the grains heats them.  A convenient visualization of this is given in fig.\,\ref{eflux} where we show the energy flux transferred from the grain to the gas as a function of time for a particular choice of the shock profile (black dashed curve).  Negative values correspond to colder grains being heated by the gas. The warmer grains sublimate their ice at a slightly higher rate than the background gas returns it, while the opposite occurs with the cooler grains.  As a result this temperature difference between different grain sizes causes the ice to slowly migrate from the warmer grains to the cooler ones while the background gas remains 100\% saturated.  The rate of migration is moderated by the number density of water molecules in the background gas, and will therefore be a sensitive function of the temperature of the gas.

For a power law size distribution with $\xi=3.5$ most of the surface area is in the smallest grains.  At the relevant temperatures ($\sim 100$~K) the Planck peak is at $\lambda\sim 30$~$\mu$m.  At these wavelengths the size parameter of Mie theory for the smallest grains is
$$x=\frac{2\pi a}{\lambda}=2\times 10^{-3}$$ 
For such small values of $x$ the cross section of the grain for both scattering and absorption is much less than its geometric cross section.  Thus despite the relatively large fraction of mass they contain, these small grains contribute very little to the grain opacity, even with their ice shells.  As the ice migrates to the larger, more efficient scatterers, it contributes more efficiently to the cross section, and the opacity increases.

Eventually, the gas temperature in the density wave becomes high enough so that the gas becomes under saturated, and the ice shells quickly sublimate from all the grains.  When the gas cools after reaching its maximum temperature, there is again a rapid condensation of the water vapor onto the silicate cores, and again, the initial thickness of the ice shell is independent of the grain radius.  Since the gas temperature is still quite high, the ice quickly migrates to the coolest grains, and as the gas cools the ice shell thickness remains unchanged.  The opacity still decreases with decreasing temperature, but this is again due to the temperature dependence of the Rosseland mean, and not to changes in the grain size distribution.

The thickness of the ice shell for the grain distribution described above is shown in Fig.\,\ref{shlthck} for the case of 0.2~AU in the density wave.  The shell thicknesses all begin at the same value, but the ice migrates to the larger, cooler, grains.  Note that it is not the largest grain in the distribution that ends up with the thickest shell.  This is fortunate, because the largest grains contribute relatively little to the effective surface area of the distribution.  Sequestering most of the ice on these grains would not allow it to contribute significantly to the opacity, and the variation of opacity with temperature would be much more moderate. 

This behavior leads to an interesting hysteresis effect.  If a grain distribution develops ice shells at a low enough temperature, the ice will migrate very slowly, or not at all.  If the shell is formed at a higher temperature and then cooled, the ice will have the opportunity to migrate to the cooler grains before being ``frozen'' in place.  The point is that the opacity of a grain distribution is not only a function of temperature at a given time, but also of thermal history of the gas and grains.  The opacity behavior with time depends on the details of this thermal history.  Below we present four examples from different regions of the density wave in more detail.  The hysteresis effect is shown for these four cases in Fig.\,\ref{opactemp} where we show the opacity as a function of temperature.  Shown for comparison are the grain opacities for these particular conditions as taken from \cite{pollack85} (dotted curve in the upper left panel) and \cite{d'alessio01} (dashed curves).  As can be seen, the agreement with published opacities is quite good before the heating event, but can differ by a factor of 2 or more after the heating event.

\subsection{0.2\,AU}

At the initial low temperatures, the difference between mass loss and mass gain is so small that the rate of ice transfer from the small grains to the large ones is barely noticeable.  As the temperature increases, however, the evaporation and condensation fluxes increase, and the mass transfer rate increases. The behavior of the opacity as a function of time for this case is shown in the upper left panel in Fig.\,\ref{opact}.  Near $t=0$ there is a sharp rise as the ice shell initially forms on the silicate cores.  Afterward there is a slow mass transfer of ice from the smaller grains to the larger ones.

As noted above, the smallest grains have a very small extinction cross section.  As a result, although they contain most of the mass of the distribution, they contribute relatively little to the opacity.  As the ice is transfered to the larger grains, the ice mass becomes more efficient at extinction, and the opacity rises.  As can be seen from the figure, the rise is sharper as the time (and hence the temperature) increases.  Eventually, however, the temperature of the gas becomes high enough that the background gas becomes under saturated.  At this point the flux off all the grains becomes greater than the flux onto the grains and all the grains lose their ice shells.  This results in a sharp drop in the opacity.  After the gas cools back down, there is an initial deposition of ice equally onto all the grains, but again the larger grains are slightly cooler than the smaller grains, and again there is a mass transfer from small to large grains.  This mass transfer is rapid because the temperatures and the fluxes are relatively high, and the ice quickly ends up on the larger grains.  As the temperature continues to drop the ice migration quickly becomes quenched.  There is a continued decline in the opacity due to the dependence of the Rosseland mean on the temperature.  The hysteresis effect in the opacity as a function of temperature is seen most clearly in the upper left panel of Fig.\,\ref{opactemp}.  Here the arrows show the direction of the opacity evolution with time.  

\subsection{0.54\,AU}

At 0.54 AU in the density wave the heating is less intense, and the temperatures are not as high, with the maximum only reaching 181\,K.  This is still high enough to evaporate the ice shells on all the grains, and the behavior is similar to the previous case.  The general shape (upper right panel in Fig.\,\ref{opact}) is the same as the previous case, as are the values of the opacity, but the times at which the grains completely evaporate or re-condense are different, because the temperature profile differs between the two cases.  The hysteresis curve for this case (upper right panel in Fig.\,\ref{opactemp}) is also similar.

\subsection{0.72\,AU}

At 0.72\,AU in the density wave, the heating is low enough so that the gas temperature never gets high enough to completely evaporate the ice shells of the largest grains.  Upon re-cooling the ice redistributes as before, but because the largest grains have retained ice from the hotter period, the overall ice distribution is different from the 0.54 AU case, and the final opacity is lower.   The opacity as a function of time for this case is shown in the lower left panel of Fig.\,\ref{opact}, and the opacity as a function of temperature in the lower left panel of Fig.\,\ref{opactemp}.

\subsection{1.32\,AU}

At 1.32\,AU in the density wave heating is weak enough so that the temperature never rises above 160\,K.  In this case the entire region where the ice shell is completely evaporated is avoided.  Instead the opacity rises monotonically to its peak value as in the first two cases and then drops slowly as the temperature decreases again (see lower right panel in Fig.\,\ref{opact}).  In the lower right panel of Fig.\,\ref{opactemp} we see that the hysteresis effect in the opacity as a function  of temperature is very similar to the first two cases except that now the drop at higher temperatures is absent since these temperatures are never reached.

It is important to note that this hysteresis effect only occurs the first time that a shock wave encounters the grains.  Once the grain distribution has experienced such a heating event, the ice is redistributed preferentially on to the colder grains.  Future heating events may remove the ice shells, but as the gas cools they will reform on the colder grains as before. This is illustrated in fig.\,\ref{repshock} where we show the opacity as a function of temperature for the 0.2\,AU case, but where the grains have encountered a second shock immediately after the first one.  The green triangles show the same case as fig.\,\ref{opactemp}, the red squares show the opacity-temperature relation for the second shock.  The profile is unchanged.  The blue diamonds show the opacity-temperature relation for the case where the grains started with an ice to olivine ratio of 2:1 by mass.  Since the ice is distributed differently among the grains in this case, the opacity differs originally as well.  Once the shock passes, the ice rearranges itself into the same distribution independent of its original configuration.

\section{Results II - Varying Dust Model Parameters}

In addition to the standard case, we have run several non-standard scenarios to test the dependence of the opacity on other parameters of the model.  These additional cases were run for a point at 0.2~AU in the density wave.  We summarize the results below.

\subsection{High-Z Case: Dust Concentration in Spiral Arms}

\cite{hagboss03a,hagboss03b} have studied the motion of solids in a disk.  As \cite{mayer07} point out, shocks can cause solids to concentrate, and thus raise the abundance of high-Z material with respect to hydrogen.  We have run a case with the same size distribution as the standard case, but with the rock and water increased by a factor of 10 over the solar value.  The result is shown in Fig.\,\ref{nonstopac}.  The panel on the upper left shows the evolution of the opacity as a function of time for this case.  Although the number of silicate cores increases by an order of magnitude, the ratio of water to silicates remains the same.  As a result, we expect that the opacity at any given time will simply be 10 times the opacity of the standard case.  A comparison with Fig.\,\ref{opact} shows that this is indeed the case.  Small differences can probably be attributed to the fact that since the background water vapor can be 10 times higher than in the standard case, there will be some differences in the rates of sublimation and condensation of the ice shell, and this will lead to small differences in opacity. 

\subsection{High Water Content in Grains}

A second case we considered was where the silicate to hydrogen ratio remains solar, but the water to silicate ratio is increased by an order of magnitude.  In this case, the ratio of water to silicate cores is 10 times that of the standard case.  So the mass of condensed water must be concentrated among fewer nuclei.  This means that the amount of available surface area corresponding to this mass will be less, and the resulting opacity will be lower.  When all of the ice vaporizes, the opacity goes down to that of the standard case, so the relative change in opacity is highest for this case.  The results are shown in the upper right panel in Fig.\,\ref{nonstopac}.

\subsection{Different Power Law For Dust Size Distribution}

Another parameter that may be varied is the power law associated with the silicate core size distribution.  The standard model used here assumed a value of $\xi=3.5$ in eq.\,\ref{eq:sizedist}.  In this case the number of small grains is much larger than the number of large grains.  In spite of this, there are still enough large grains available so that when the ice shells on the smaller grains evaporate, there are enough large grains to take up the condensate.  As a result, we see the hysteresis effect discussed above.  This is illustrated by the red curves in the two bottom panels of Fig.\,\ref{nonstopac}.  

For $\xi=1.5$ (blue curves in Fig.\,\ref{nonstopac}), the number of small grains is not so much larger than that of the large grains.  As a result, when the smaller grains lose their ice shells, the change in the ice shell radius of the larger grains is more moderate, and the opacity doesn't increase that much.  The hysteresis effect is therefore smaller, as seen in the figure.

For $\xi=5.5$ (green curves in Fig.\,\ref{nonstopac}), there are many more small grains, and the ice is initially divided mostly among these small grains.  As a result, the ice shells are quite thin.  In addition, the opacity is mostly due to the small grains, since they are by far the most abundant.  Even though the ice migrates to the larger grains as before, the small grains dominate the opacity, so that hysteresis effect is small and the total opacity actually decreases as the ice migrates to the larger grains.  In order to save time these last three cases were run with a maximum grain size of 10~$\mu$m instead of 1~mm as in the standard case.  As can be seen by comparing the case for $\xi=3.5$ in the figure with the corresponding case in Fig.\,\ref{opact}, the opacity here is higher because more of the grain mass is found in the more efficient scatterers.

\subsection{Low Opacity Regions and the Effect of Stellar Irradiation}

All the cases described above were run for the case of high opacity because we have considered shock profiles at the midplane of the disk.  In this case the radiative heating is due to the surrounding gas.  Since this gas is at a temperature of the order of 100\,K, the radiation it emits will be mostly in the infrared.  Well above the disk plane, however, the opacity drops, and the visible radiation from the star can penetrate into the disk.  Because the radiation from the star is concentrated at much shorter wavelengths, the size parameter of the grains with respect to this wavelength increases, and their efficiency for scattering and absorption changes.  In addition, water ice is essentially transparent in the visible, so that the albedo of the grains changes as well.  The magnitude of any changes due to this effect are reduced by the fact that these density waves only appear at 10\,AU or more from the star.  

The total optical depth in the inter-arm region is of the order of 100, so the region where the optical depth is of order unity comprises only a few percent of the total mass in a column.  In the arm itself, the total optical depth is a factor of 10 higher, so the low optical depth region is an even smaller fraction of the total mass.  Nonetheless, it is of interest to see how the evolutionary behavior of the ice grains is affected by the addition of stellar radiation.

Figure\,\ref{lowoptt} shows the opacity as a function of temperature at 0.2\,AU in the arm (the same case as shown in the upper left panel of Fig.\,\ref{opactemp}).  Here, however, we have added the stellar radiation due to a star at a distance of 10\,AU.  The parameters of the star (radius and surface temperature) are the same as for our sun.  The blue curve shows the case where the gas density is the same as the density for the optically thick case shown in the upper left panel of Fig.~\ref{opactemp}, so that the only additional effect is the added stellar radiation.  The grains in the first bin have a silicate core of radius 0.022\,$\mu$m.  At low temperatures, with an ice shell, they grow to a radius of $\sim 0.2\,\mu$m but most of this additional radius is composed of nearly transparent ice.  As a result, their extinction cross section, even in the visible is small, and the additional stellar heating does not affect their temperature appreciably.  The grains in the larger bins absorb a larger percentage of the stellar heating, however, and reach higher temperatures.  As a result, the ice migrates to the smallest grains, rather than to the larger ones as in the high opacity case.  At the IR wavelengths that are relevant to the Rosseland mean, the smallest grains do not contribute much to the opacity, and the opacity is about half what it is in the high optical depth case.

Since the optical depth depends on the gas density, we would expect that this density is much lower in the low optical depth region.  The red curve in Fig.\,\ref{lowoptt} for a gas density $10^{-4}$ of midplane is therefore more appropriate.  Here, because of low gas density, cooling by contact with the gas is inefficient.  In this case the grains reach much higher temperatures than the gas, and the coolest grains are those in bin 4, with core radii of 22~$\mu$m.  Even these grains are too hot to retain ice shells until the gas temperature drops to below around 125~K and the opacity starts to rise.  Even so, the opacity is still much lower than the corresponding value at midplane.

\section{Results III - The Case of Composite Dust Grains}

So far we have considered the grains to consist of a silicate core with an ice mantle, but this is likely a very simplified model of the actual structure of grains.  In particular, suppose the grains consist of silicate micrograins embedded in an ice matrix.  In this case, the effect on opacity can be much more dramatic.

Consider first a case similar to that investigated above.  If the grains are assumed to be composed of a mixture of rock and ice, and we assume that the ice to rock ratio, by mass, is 2:1, then $M_i=2M_r$, where $M_i$ is the ice mass and $M_r$ is the rock mass.  This means that the mass fraction of ice, $X_i=2/3$ and the mass fraction of rock is $X_r=1/3$.  If we take the bulk density of ice to be $\rho_i=1$~g~cm$^{-3}$ and the bulk density of the silicate to be $\rho_r=3$~g~cm$^{-3}$, the density of the mixture, $\rho$ is 
$$\frac{1}{\rho}=\frac{X_i}{\rho_i}+\frac{X_r}{\rho_r}$$ and $\rho=1.29$~g~cm$^{-3}$.  For a given mass of grain material, $M$, the number of grains, $N$ is given by 
\begin{equation}
M=M_i+M_r=\frac{4\pi}{3}Na^3\rho
\end{equation}
and 
\begin{equation}
N=\frac{3M}{4\pi a^3\rho}
\end{equation}

To a zeroth approximation, we can neglect the differences in refractive index between water and rock, and assume that they all have a refractive index of $m=1.5+0.1i$.  Then $$\frac{m^2-1}{m^2+2}=.296+.05i$$ and, if the size parameter of the grain is given by
$$x=\frac{2\pi}{\lambda}$$ where $\lambda$ is the wavelength of the incident radiation, the efficiency factor for extinction for $x\leq 1$ can be well approximated by \citep{hulst57}
\begin{equation}
Q_e=Q_0x^4=x^4\Re\left[\frac{8}{3}\left(\frac{m^2-1}{m^2+2}\right)^2\right]
\end{equation}
where the symbol $\Re$ refers to the real part of the expression in the brackets.  For our case this gives $Q_e=0.0851x^4$.  For a given mass of grains, therefore, the opacity will be proportional to 
\begin{equation}
N\pi a^2Q_e=\frac{3MQ_e}{4a\rho}=Q_0\frac{3M}{4a\rho}x^4
\end{equation}

If the ice now evaporates, the new grain mass will be $M^*=X_rM$.  The new grain bulk density will be $\rho^*=\rho_r$, and the new grain radius will be
$$a^*=a\left(\frac{X_r\rho}{\rho_r}\right)^{1/3}$$
If the original opacity is
$$
\kappa=\kappa_0NQ_e\pi a^2
$$
the new opacity will be
\begin{equation}
\kappa^*=\kappa_0Q_0\frac{3M^*x^{*4}}{4a^*\rho^*}=\kappa\left(\frac{X_r\rho}{\rho_r}\right)^2
\end{equation}
For the parameters we have chosen, 
$$\left(\frac{X_r\rho}{\rho_r}\right)\approx 0.15$$
so the opacity should drop to around 0.02 of its original value.  In fact, in the full model we see a smaller change, around 0.1, but this is partially due to the fact that we are using the same refractive index for water, rock, and the water + rock mixture.  Since ice is more transparent than rock, and in this simple model is assumed to have the same extinction properties as rock, the removal of ice will overestimate the change in opacity.  More important, however, is the fact that the higher opacity occurs at lower temperature, where the peak of the Planck function is at longer wavelengths.  In going from 130~K to 220~K the wavelength decreases by a factor of 0.6, so $x$ increases by the same factor, and $x^4$ increases by a factor of around 8, so the overall change is the product of these two factors $\approx .16$, which, allowing for the previously mentioned refractive index effect, is about right.  So, to zeroth order, the opacity change for grains with $x\leq 1$ is
\begin{equation}\label{vap}
\kappa^*=\kappa\left(\frac{X_r\rho}{\rho_r}\right)^2\left(\frac{T^*}{T}\right)^4
\end{equation}
where $T^*$ is the high temperature at the place where the ice has evaporated, and $T$ is the low temperature at the place where the ice is in the grain.

Now suppose that the grains are actually composed of a conglomerate of micrograins held together in an ice matrix.  As the ice evaporates, the micrograins are released.  Suppose that each large grain of radius $a$ contains micrograins of radius $a'$.  Again the ratio of ice to rock is 2:1.  Now when the ice disappears, the micrograins are released and the relevant parameters change as follows: $M^*=X_rM$, and $\rho^*=\rho_r$ as before, but now $a'=a$ and $$x^*=x\frac{a'}{a}$$ $$N^*=\left(\frac{a}{a'}\right)^3\frac{X_r\rho}{\rho_r}N$$  As a result
\begin{equation}\label{micro}
\kappa^*=\kappa \frac{X_r\rho}{\rho_r}\left(\frac{a'}{a}\right)^3\left(\frac{T^*}{T}\right)^4
\end{equation}

Note that when $a'=a^*$ for the simple ice shell evaporation, eq.\,\ref{micro} reduces to eq.\,\ref{vap} as it should.  If $a\sim 1\mu$m and $a'\sim 0.1\mu$m, for example, the reduction in opacity could be very dramatic.

\section{Discussion}

Using the time-dependent spiral shock profiles from a 3-D hydrodynamical simulation of a gravitationally unstable disk we have calculated the effects that such shocks have on the opacity of the gas. Our results show that opacity variations of about an order of magnitude can take place on timescales of several months to years. Such timescales are much shorter than the orbital time, which is larger than 10 years at the distances of the shocks in the simulations (10\,-\,20\,AU from the star). This means that the cooling time in the disk will change on timescales much shorter than the orbital time. In particular, when the opacity drops by nearly an order of magnitude on such short timescales it means that the cooling time will also decrease proportionally fast on such short timescales  \citep[see also][]{cossins10}.  The cooling disk may not be able to self-regulate on such short timescales and therefore the likelihood of fragmentation should be substantially increased.  Likewise, the temperature increase caused by the shock provides the right conditions to produce a rapid increase in molecular weight within the spiral shock, supporting the claim of \cite{mayer07}, as long as rapid migration of solids toward the arms can occur simultaneously and increase the dust-to-gas ratio. 

The water vapor that is released by the grains will increase the mean molecular weight of the gas.  As \cite{mayer07} point out, this will make the disk more unstable.  If the molecular weight of the background gas is $\mu_x$ and this gas is a mass fraction $X$ of the total, then introducing a gas of molecular weight $\mu_z$ and mass fraction $1-X$ will increase the mean molecular weight, $\overline{\mu}$, to
\begin{equation}\label{meanmu}
\overline{\mu}=\mu_x\left[1+\frac{\mu_z-\mu_x}{\mu_z}\left(\frac{1-X}{X}\right)\right]
\end{equation}

For solar composition gas, $\mu_x\approx 2.4$, while $\mu_z=18$ for water vapor.  For the baseline model we investigated the water vapor mass fraction is $9\times 10^{-3}$ so that $\overline{\mu}$ would increase by only 1\%.  But if grains are concentrated by the spiral arms and the mass fraction of water vapor increase by an order of magnitude, $\overline{\mu}$ can increase from 2.4 to 2.6.  This is large enough to have noticeable effects on the stability of the disk \citep{mayer07}.

No published hydrodynamical calculations exist that can show the simultaneous effect of both variations. By combining the results of \cite{mayer07} and \cite{merubate10}, one can expect that the formation of giant planets in the inner disk would be much more likely with a concurrent rapid drop of opacity and increase in molecular weight, probably irrespective of whether convection occurs or not [see also \cite{jongam03}].

\cite{rudpol91} have pointed out that if the opacity, $\kappa$, has a temperature dependence of the form $$\kappa\sim T^{\beta}$$ then a gas with an adiabatic exponent of 1.45-1.5 (which is a good approximation in the temperature range of the shock at peak
strength, between 150 and 240 K, for solar metallicity, see \cite{boleyetal07}) will satisfy the Schwarzschild criterion for convective instability if $\beta>0.5-1$. The rising parts of the dashed curves in Fig.\,\ref{opactemp} have slopes corresponding to $\beta=0.5$.  As it can be seen from the figure, when the shock ramps up, in the region with T= 150\,K where the ice quickly migrates to the largest cores, the value of $\beta$ is temporarily $ > 1$, namely it is above the critical value expected at this temperature, for which $\gamma = 1.5$ according to \citep{boleyetal07}. The exact value of $\beta$ depends on the location within the shock (i.e. it varies across different annuli because the opacity-temperature relation varies), but it is always high enough
to allow, in principle, the development of convection. Later on, as the temperature increases and then decreases again, the adiabatic index also varies, thus changing the critical threshold, and $\beta$ becomes sub-critical. The same is true for a second passage of the shock since the opacity has also changed in the meantime. Therefore, it appears that convection could be triggered initially but would be short-lived. However, since convection itself could change the density and temperature distribution through the shocking region, only a dynamical simulation which includes our coated grain model will be able to 
ascertain the thermodynamical evolution of the gas phase. Finally, in this discussion we have not considered how the conditions might change as a function of the altitude from the midplane since we have used azimuthally averaged quantities, but going beyond this will once again require a dynamical (3-D) simulations since convection, if it starts, is expected to change the physical conditions particularly across the vertical extent of the disk.

Another possible effect could arise because molecular weight variations can in principle damp the convective instability.  Indeed, since vertical convection requires $ds/dz < 0$, where $s$ is the specific entropy, and, ignoring adiabatic index gradients, $ds/dz = d \ln T/dz - (\gamma-1) d ln \rho/dz - d ln \mu/dz$ for an ideal gas, one sees that a vertical gradient in the molecular weight
can have a stabilizing influence on convection. Therefore, it will be important to understand whether or not significant $\mu$ gradients would develop once $\mu$ can be self-consistently calculated in simulations including both gas and dust. The important point to remember is that, for convection to be relevant for disk fragmentation, what matters is that it affects the cooling of the disk at midplane, and not necessarily across the entire vertical extent of the disk. Therefore, as long as molecular weight gradients are negligible near the disk midplane convection will not be affected. In practice, one will need to show that there is a large enough region around the midplane in which heat can be redistributed rapidly by convection but within which the molecular weight is increased nearly uniformly following dust grain collection and sublimation through the shock.

It is tempting to speculate that the gas becomes convectively unstable at this point, and this convection allows the gas to cool more quickly.  If this is indeed the case, then the increase in opacity would be offset by convective cooling.  Note that this high $\beta$ occurs only during the heating of the gas, and $\beta$ is negative during the cooling stage.

Certainly the influence of ice coatings on grains needs to be
investigated directly with numerical simulations since these effects are not necessarily additive; for example, one can imagine that an increasing molecular weight would affect the temperature evolution. In addition, the opacity variation across the shock would also be affected if the timescales of the two processes are even slightly uneven.

In general, the present work shows the importance of modeling grain chemistry within the spiral shocks of self-gravitating protoplanetary disks. While at $50\,{\rm AU}  < R < 200$\,AU fragmentation in disks loaded by high accretion rates from the molecular envelope is an almost unavoidable event which depends little on the details of the gas and dust chemistry because the cooling time is expected to be short \citep{boley09,boleyetal10,rafikov09,vorobyov10b}, fragmentation in the inner disk ($< 30$\,AU) is going to depend very strongly on thermodynamics and therefore on the properties and response of dust grains that affect the cooling time via the opacity.  Other thermodynamical factors, such as the effective adiabatic index \citep{boleydur08} and the molecular weight \citep{mayer07}, are also bound to play an important role because they will change the temperature evolution across the shock.

The models that we describe in this paper are not dynamical; the grains do not move in response to the gas, nor do they coagulate and change their size distribution.  Recent calculations \citep{boleydur10} show that grains significantly above interstellar size, can migrate toward spiral arms, increasing the dust-to-gas ratio and even affecting fragmentation itself. In our calculations we have assumed a size distribution appropriate for interstellar grains, and no grain growth (other than the growth of an ice shell), so that the assumption of coupling between gas and dust is satisfied. In the spiral shock the dust-to-gas ratio might go up by a factor of 10 if most grains have already grown to cm-size grains.  This could open a new avenue for forming planetesimals rapidly via gravitational instability of the dust layer \citep{riceetal06}. Dust accumulation in spiral shocks might also be an additional effect promoting fragmentation in the gas phase \citep{boleydur10}, concurrent with those that we have just described.

The results of our paper show that the detailed modeling of the response of the dust grains to spiral shocks has an important effect on the evolution of the opacity, one of the crucial parameters controlling the development and outcome of gravitational instability in massive disks. In their present form our results should be regarded as a "proof of concept" since the temperature 
evolution of the shocks that we have assumed in the paper has not been recomputed using our dust model, but rather stems from the \citep{d'alessio01} model adopted in the simulations. Incorporating the dust model that we have described in this paper directly in our three-dimensional hydrodynamical simulations of disks will be the first and most important next step. The stronger and faster opacity changes that we have found might produce fragmentation in a disk model that does not fragment with the standard opacity models used in \cite{mayer07} as well as in other published work. Likewise, since grain growth might have already taken place by the time the disk becomes spirally unstable, we will need to extend our model to larger grains, with sizes comparable to those that can accumulate efficiently within spiral shocks.

\begin{figure}[ht]
\centerline{\includegraphics [width=12cm]{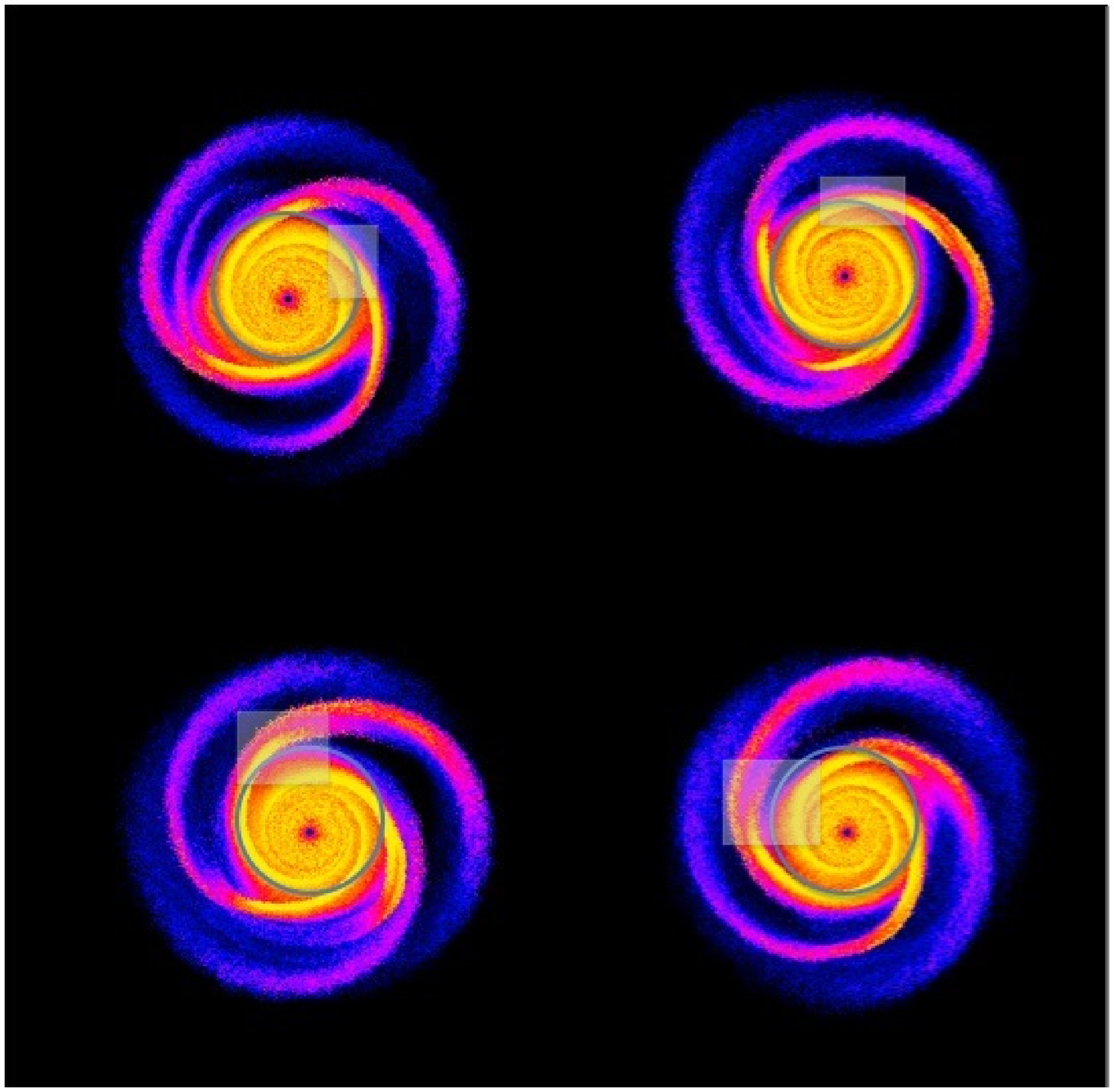}}
\caption{Color-coded temperature maps of the hydrodynamical disk 
simulation. The temperature ranges from 30\,K (dark blue) to 250\,K (yellow).
The $m=3$ mode is evident, with three strong spiral shocks developing in 
the region located between 12 and 15 AU from the central star (shown as a 
dark spot at the center). The disk is approximately 35 AU in radius, with 
the outer boundary naturally changing in time as a result of outward 
angular momentum transport from spiral density waves. The box is identified in a Lagrangian fashion by tracking the particles initially within the shock,and therefore its size changes in time (see text for a detailed description). Initially the box has an azimuthal extent of about 4\,AU and a height of 0.15\, AU.  The four panels, equally spaced in time, cover about 20 years of evolution 
(from upper left to bottom right) before and after one of the spiral shocks reaches maximum amplitude (which happens in the bottom left panel), approximately corresponding to the time spanned by the profiles in Fig.\,\ref{shock}.  The maximum amplitude corresponds to the temperature maximum shown in Fig.\,\ref{shock}.  The azimuthally averaged profiles in Fig.\,\ref{shock} have been computed setting the location of the overlaid circle to $r=0$ and then considering increasingly larger annuli (with widths indicated as in Fig.\,\ref{shock}) within the boundaries of the box.} 
\end{figure}\label{shockfig1}

\begin{figure}
\centerline{\includegraphics[width=12cm]{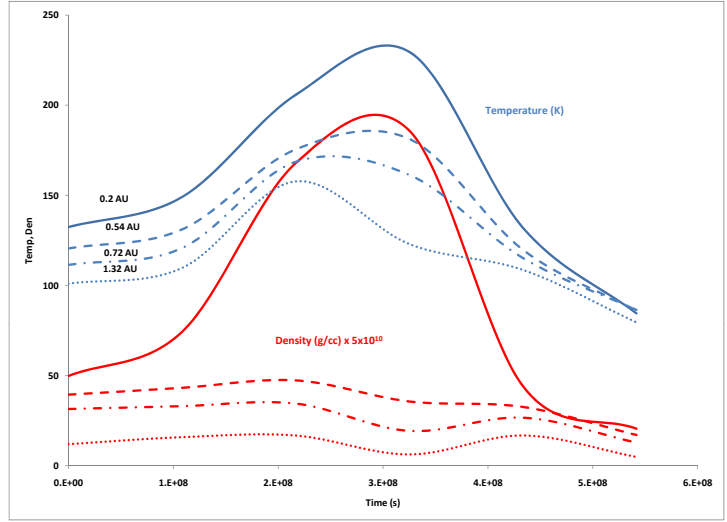}}
\caption{Azimuthally averaged gas temperature (blue curves) and gas density ($\times 10^5$) (red curves) for three regions in the spiral shock. measured at the disk midplane.  The three regions correspond to annuli at $r= 0.2 AU$ (solid curve), 0.54 AU (dashed curve) and 1.3 AU (dotted curve), where $r=0$ is defined as the radius of the reference circle that approximately follows the shock peak amplitude, as shown in Figure 1.} \label{shock}
\end{figure}

\begin{figure}[p]
\centerline{\includegraphics[width=10cm]{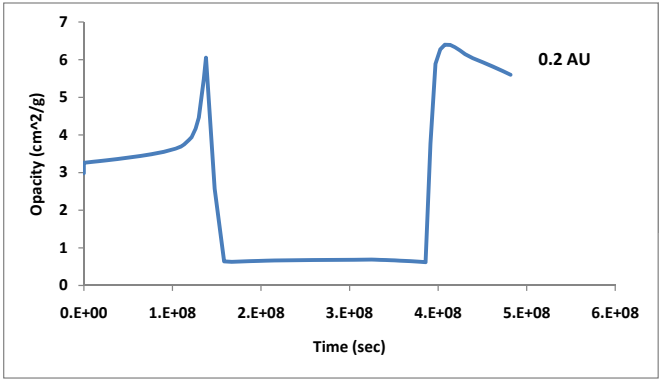}\includegraphics[width=10cm]{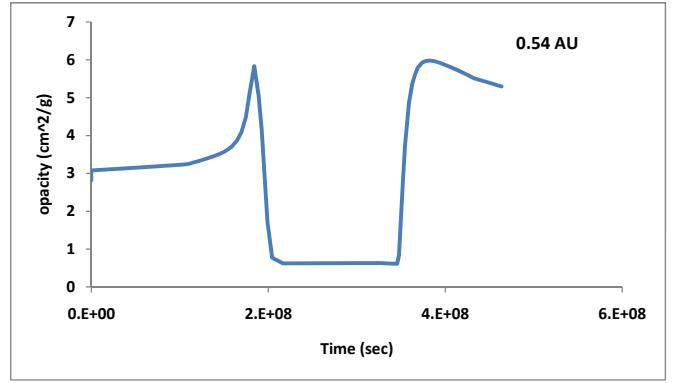}}
\centerline{\includegraphics[width=10cm]{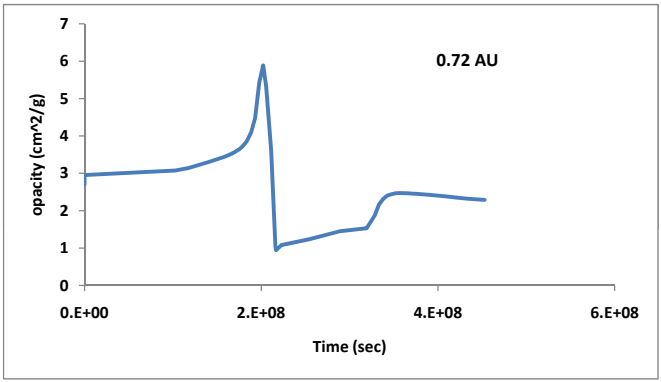}\includegraphics[width=10cm]{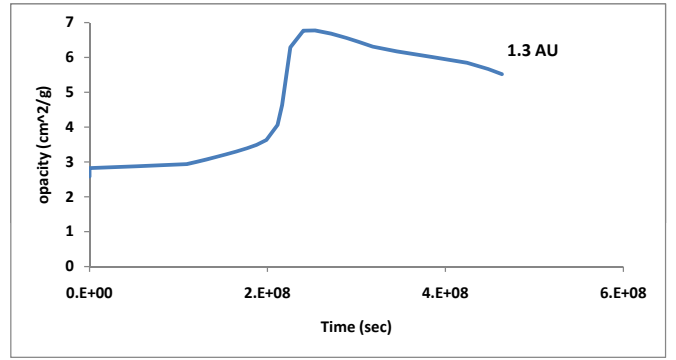}}
\caption{Opacity as a function of time for the annuli at 0.2 AU (upper left), 0.54AU (upper right), 0.72AU (lower left) and 1.3AU (lower right) in the density wave.} \label{opact}
\end{figure}

\begin{figure}
\centerline{\includegraphics[width=20cm]{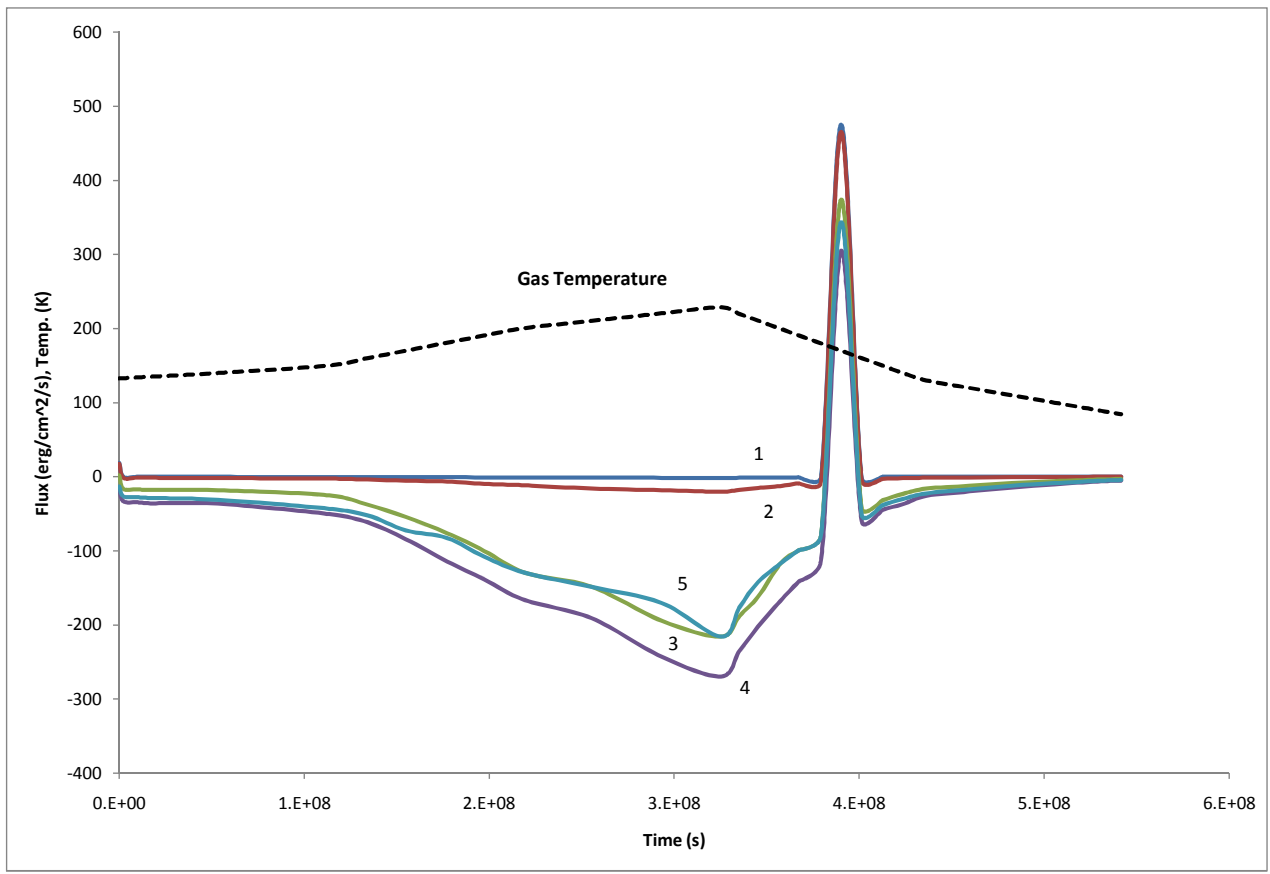}}
\caption{Energy flux from the grain to the gas as a function of time for the 5 grain sizes considered.  Negative values indicate that the grain is colder than the gas.  The numbers next to each curve refer to the grain size bin number corresponding to that curve.  The gas temperature is shown as a dashed curve.  The peak flux near $4\times 10^8$\,s is due to water vapor recondensing on the grains and heating them.  This corresponds to a gas temperature of 165\,K, with the grains about 0.3\,K hotter.}
\label{eflux}
\end{figure}

\begin{figure}[p]
\centerline{\includegraphics[width=15cm]{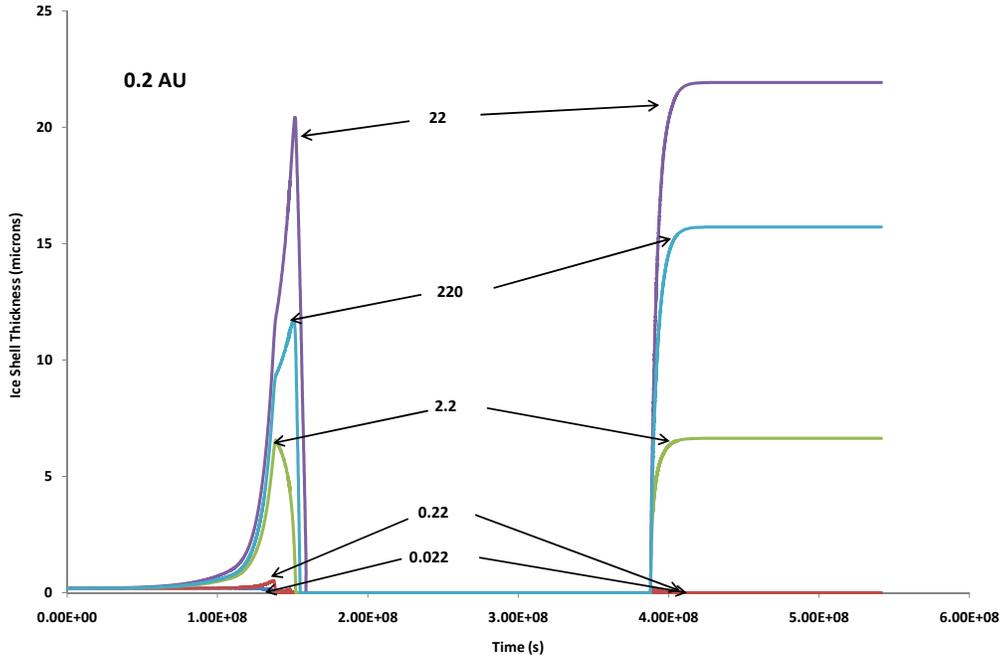}}
\caption{Shell thickness as a function of time for the 0.2\,AU annulus in the density wave. The numbers next to the curves give the core radius in microns.  Note that the thickest shell does not surround the largest core.}
\label{shlthck}
\end{figure}

\begin{figure}[p]
\centerline{\includegraphics[width=8cm]{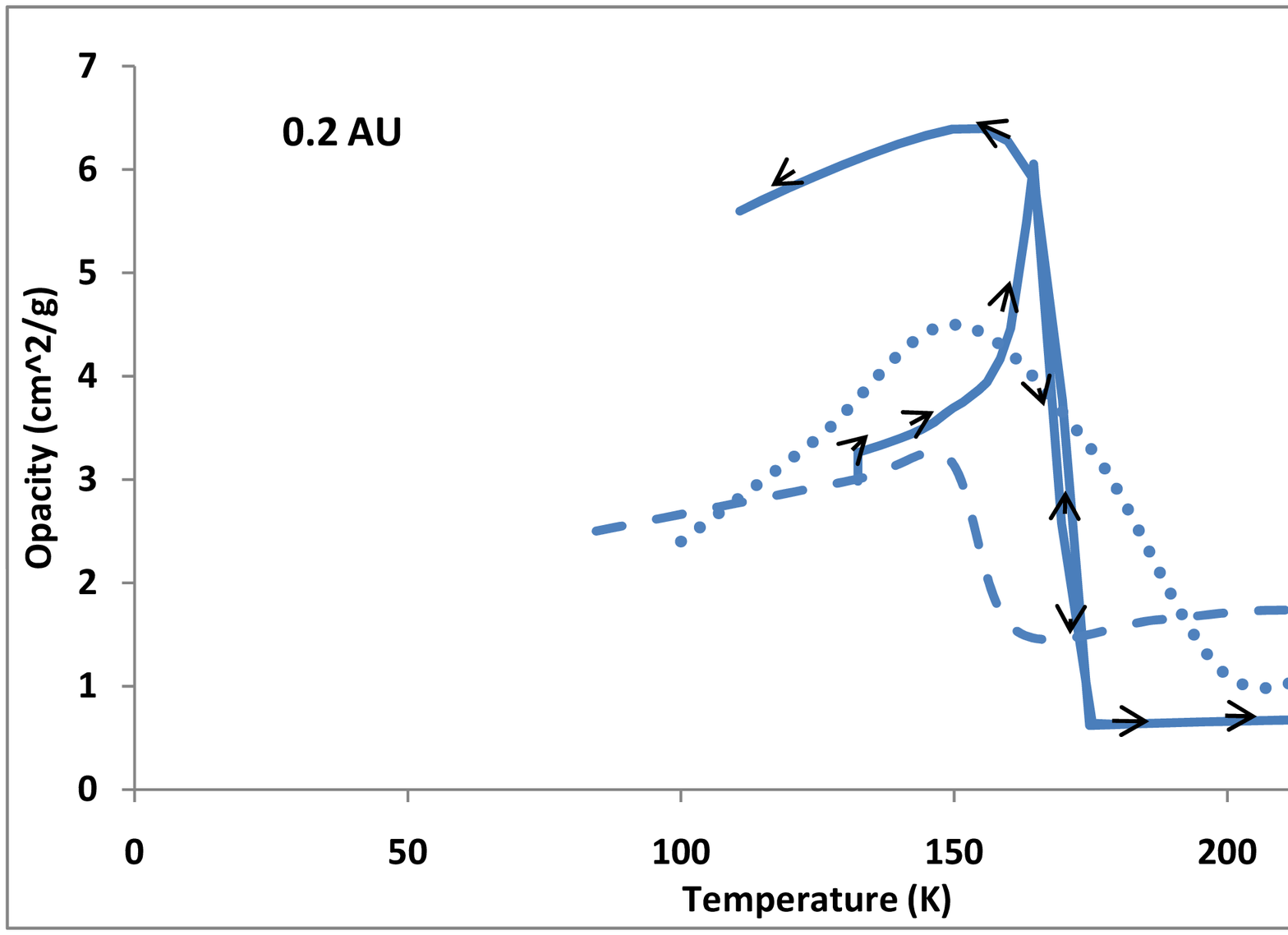}\includegraphics[width=8cm]{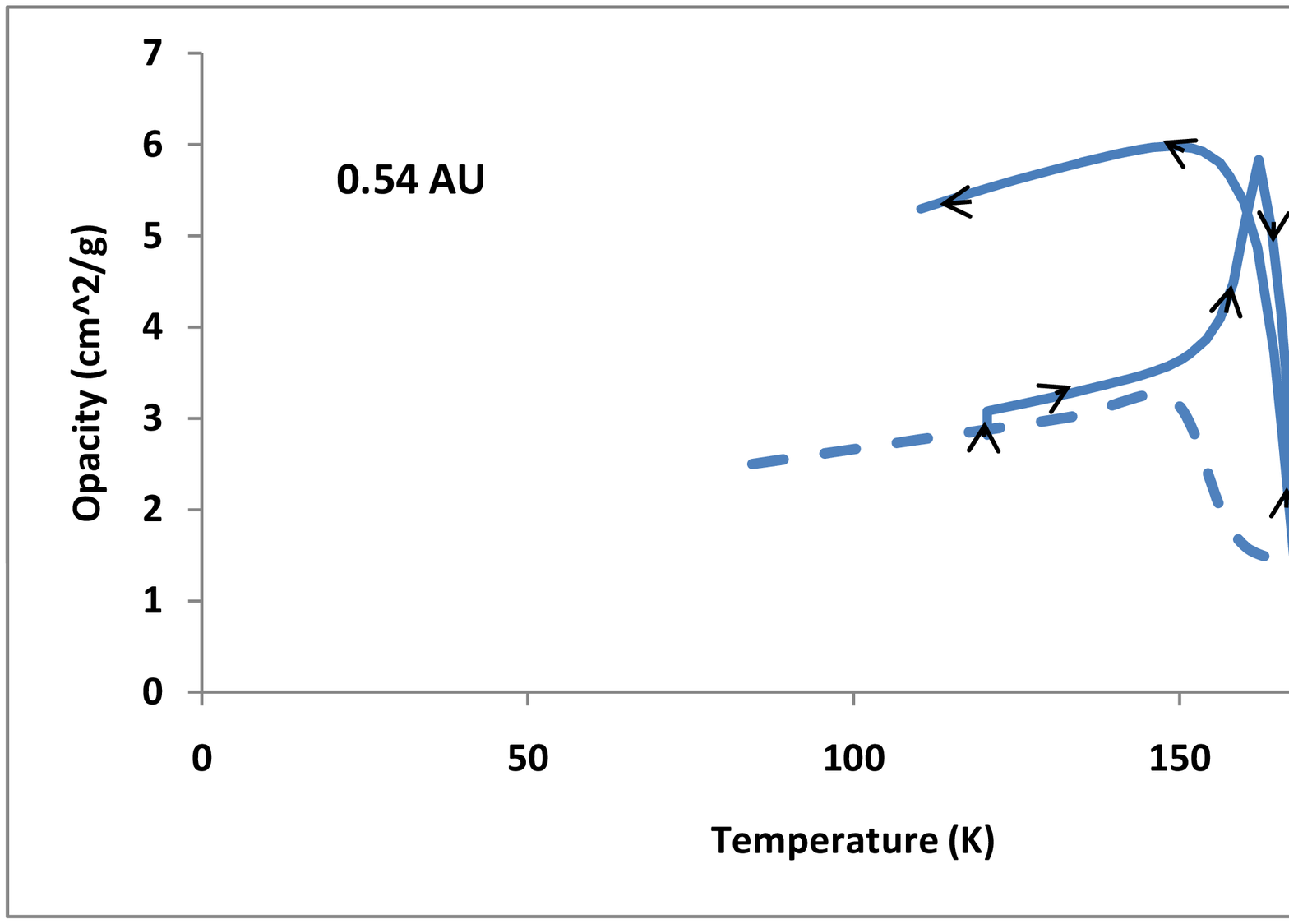}}
\centerline{\includegraphics[width=8cm]{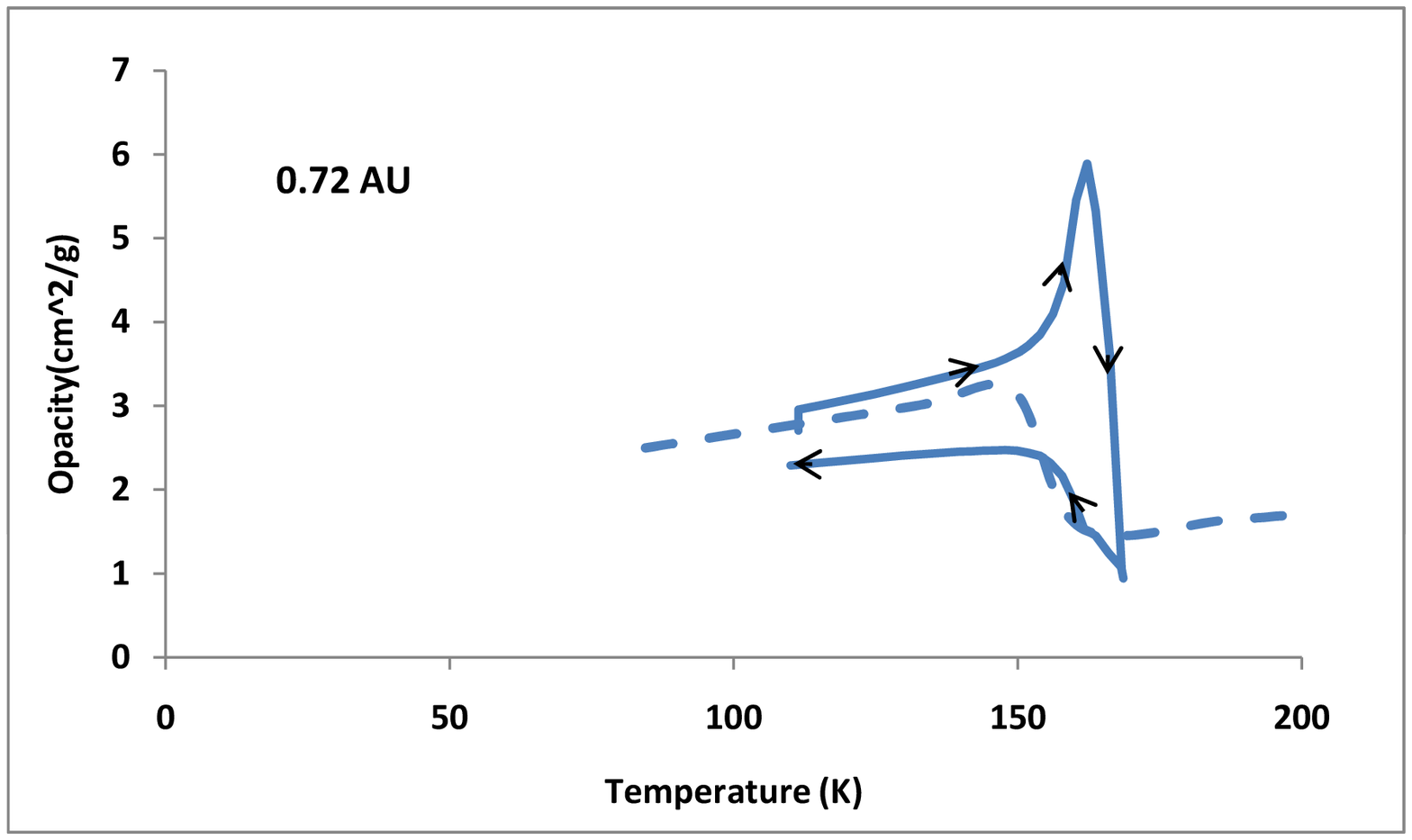}\includegraphics[width=8cm]{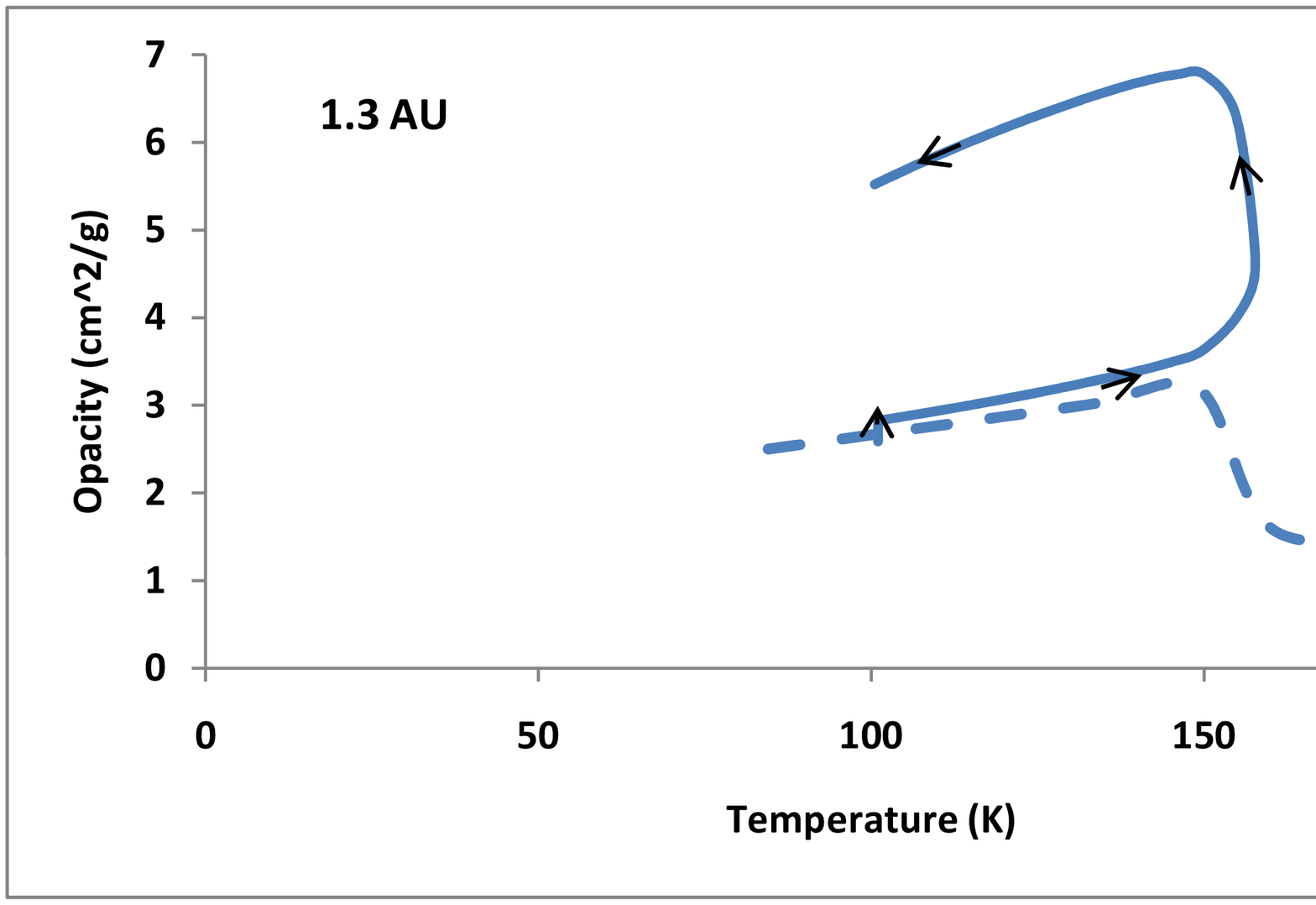}}
\caption{Opacity as a function of temperature for the 0.2 AU (upper left), 0.54AU (upper right), 0.72AU (lower left) and 1.3AU (lower right) annuli
in the density wave.  The black arrows show the direction of the opacity change as a function of time.  The dotted curve in the upper left hand panel shows the opacity of \cite{pollack85} while the dashed curves in all the panels show the opacities of \cite{d'alessio01}.}
\label{opactemp}
\end{figure}

\begin{figure}
\centerline{\includegraphics[width=20cm]{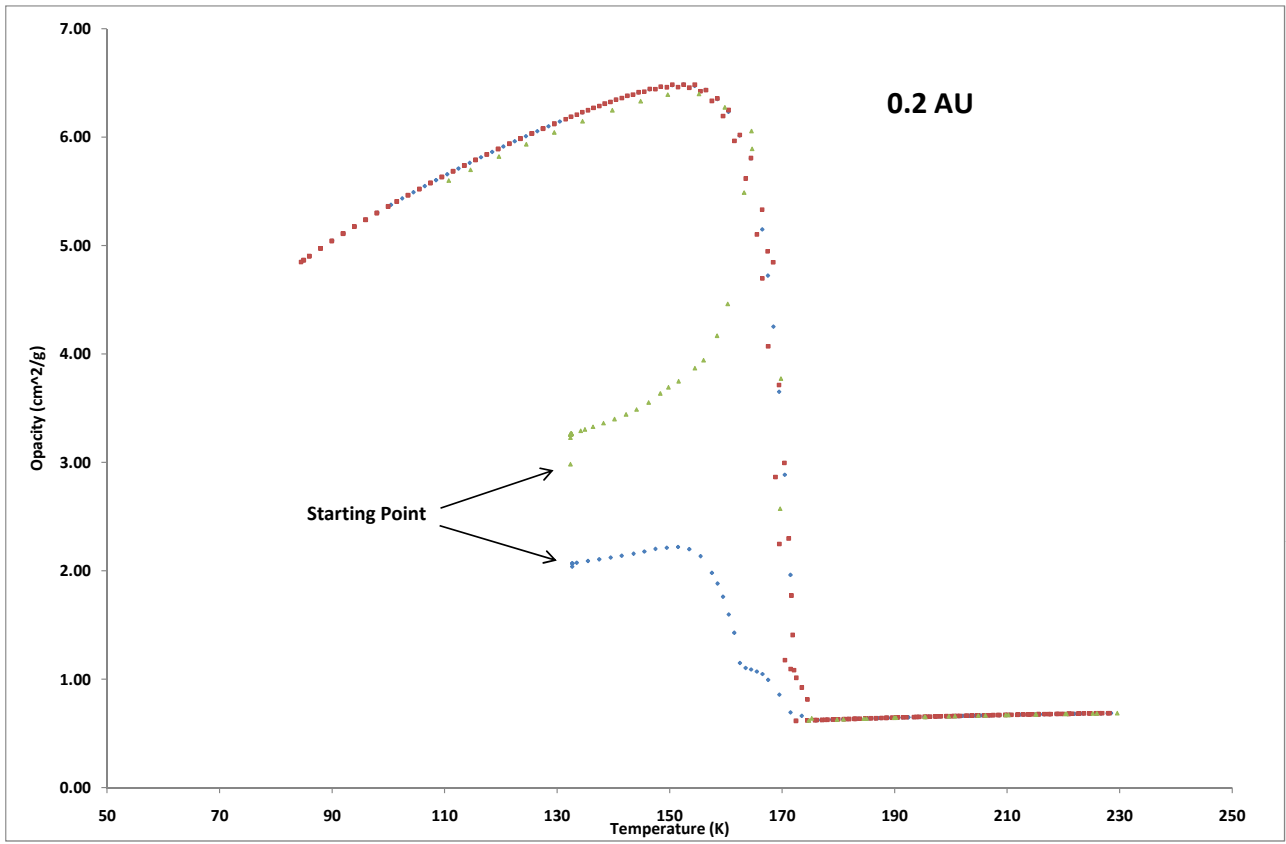}}
\caption{Opacity as a function of temperature for the 0.2\,AU case. The green triangles show the case discussed earlier (no ice on the olivine cores at $t=0$).  The blue diamonds show the same case but with an ice to olivine ratio of 2:1 at $t=0$.  Because the original ice arrangement is different the opacity is different at the starting point ($t=0$).  Once the shock heats the grains enough for the ice to migrate, it quickly rearranges itself in the same way, independent of the original distribution.  After a second shock (red squares) the opacity - temperature relation remains unchanged.}
\label{repshock}
\end{figure}

\begin{figure}
\centerline{\includegraphics[width=8cm]{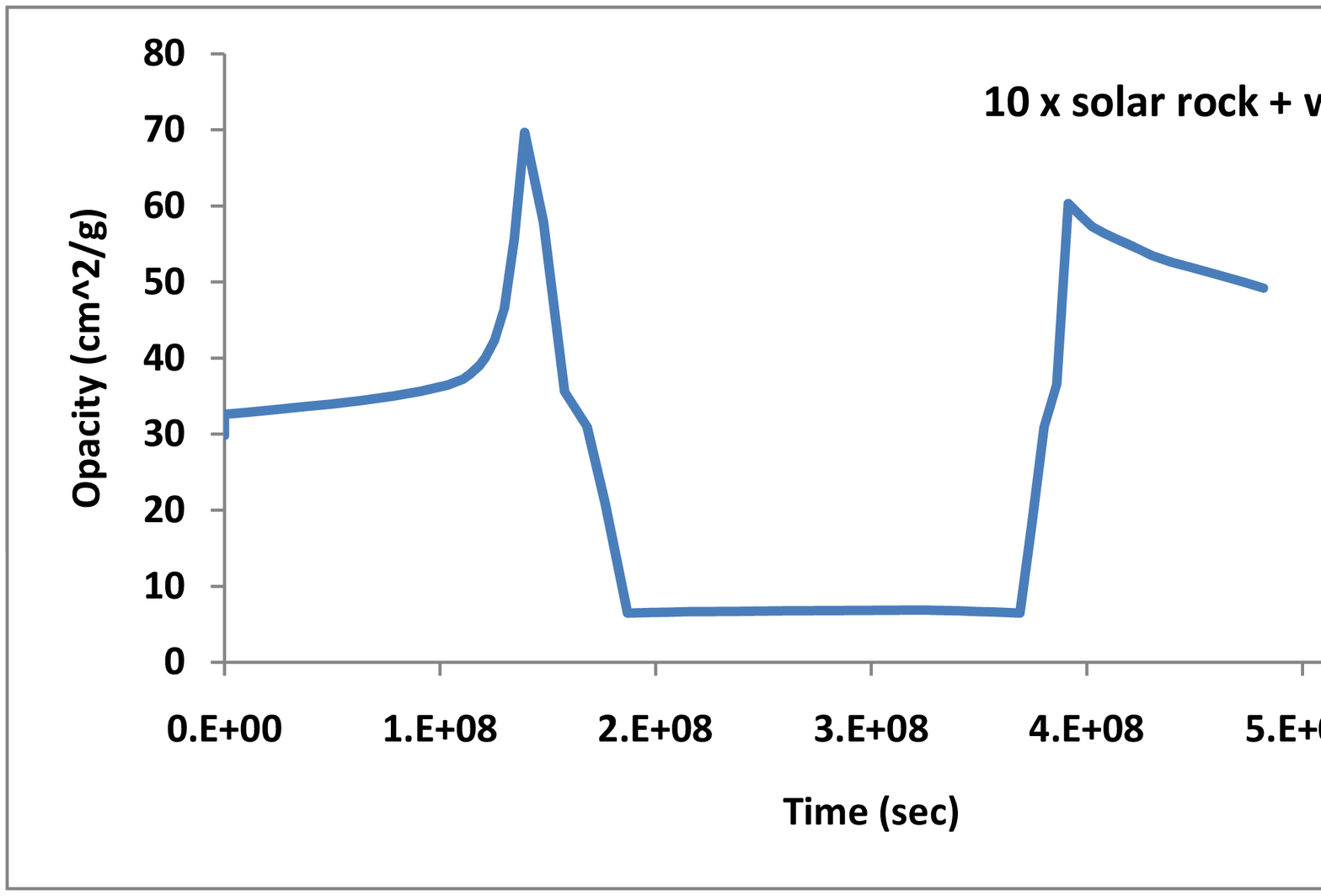}\includegraphics[width=8cm]{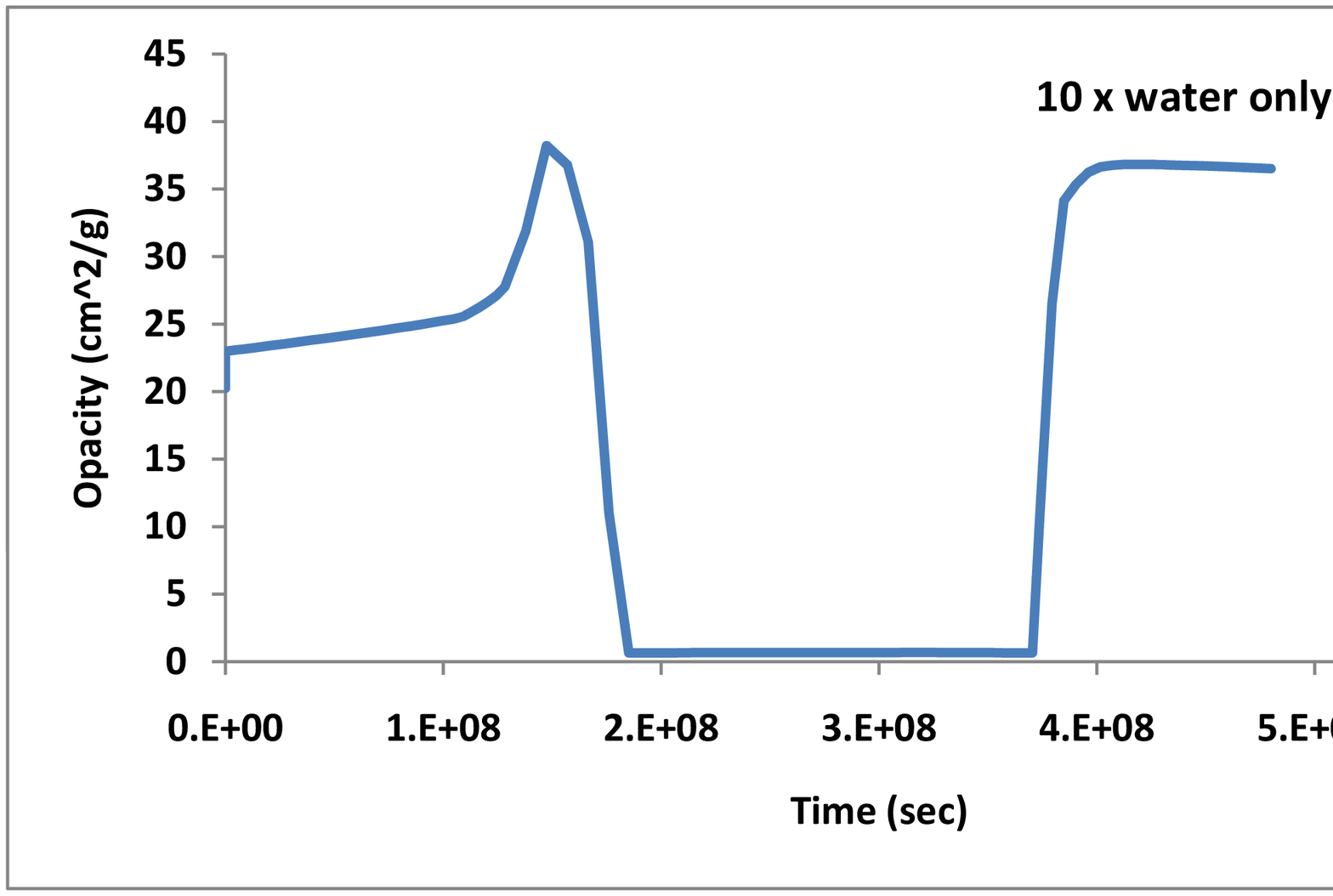}}
\centerline{\includegraphics[width=8cm]{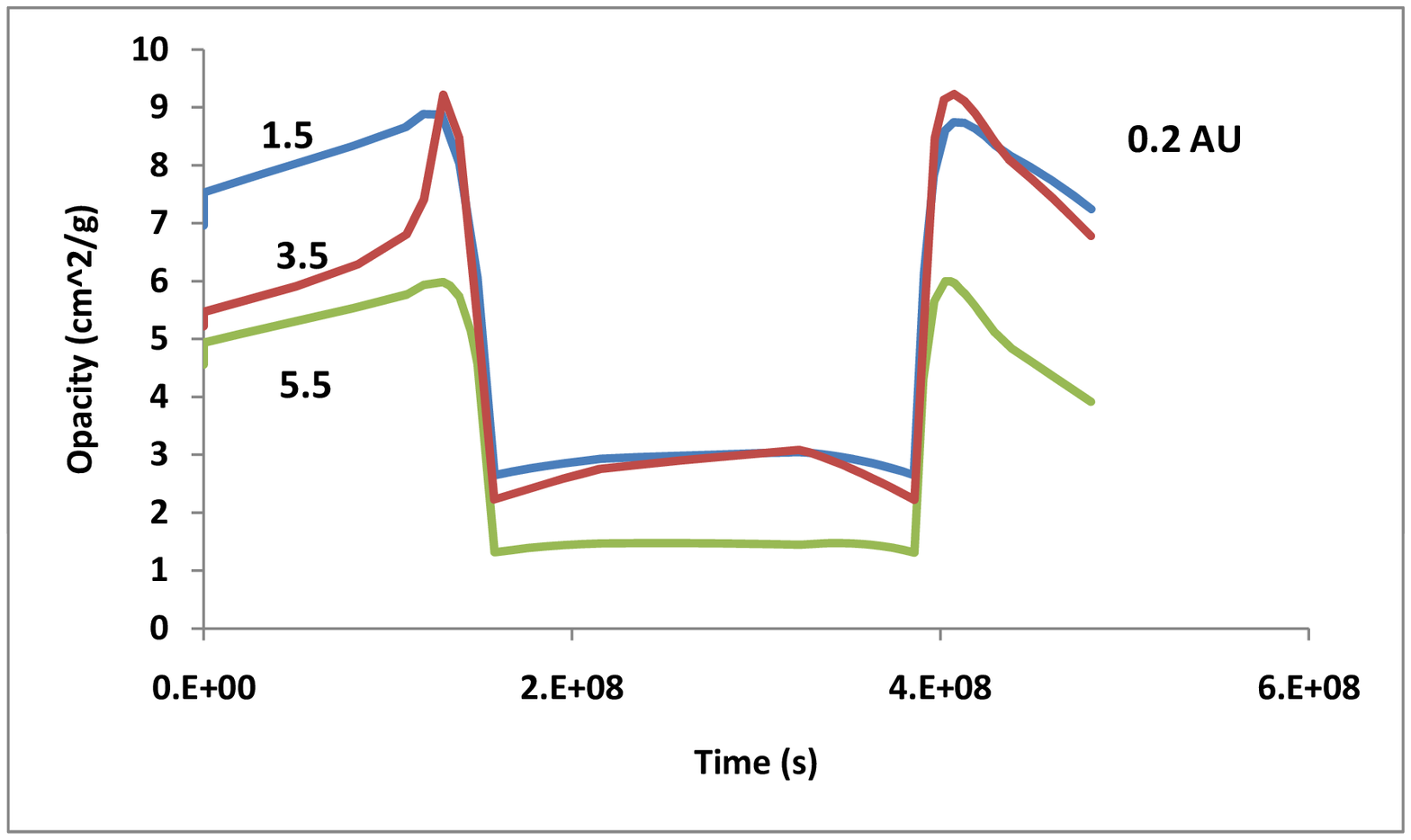}\includegraphics[width=8cm]{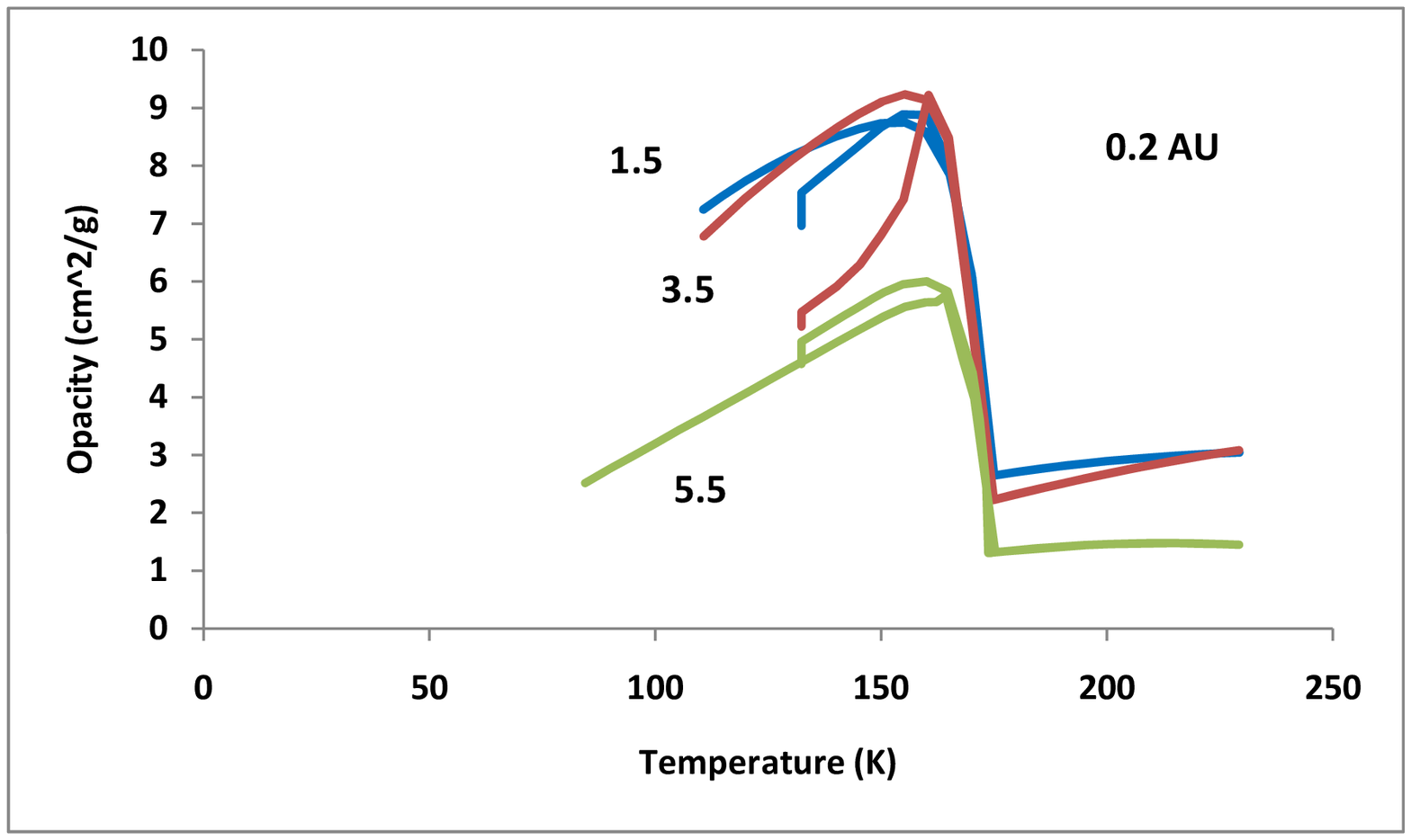}}
\caption{Opacity as a function of temperature for the 0.2\,AU annulus in the shock but where the abundances are 10 times the solar value (upper left), and for where only the water is 10 times the solar value(upper right).  Also shown are solar abundances with 3 values of $\xi$: 1.5 (blue curve), 3.5 (red curve), and 5.5 (green curve).  The opacity as a function of time is shown in the lower left hand panel and the opacity as a function of temperature is shown in the lower right hand panel.}
\label{nonstopac}
\end{figure}

\begin{figure}
\centerline{\includegraphics[width=14cm]{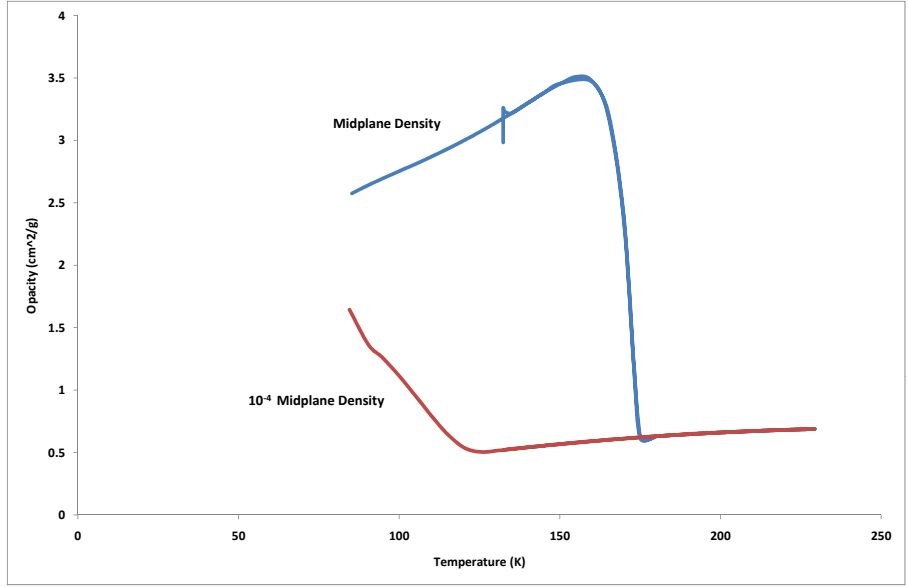}}
\caption{Opacity as a function of temperature for the case of 0.2 AU in the shock but where the optical depth to the star is essentially zero.  The blue curve is for a gas density equal to the gas density at midplane, and the red curve is for $10^{-4}$ of the midplane density.}
\label{lowoptt}
\end{figure}

\bibliographystyle{apj}
\bibliography{paper} 

\begin{thebibliography}{31}
\expandafter\ifx\csname natexlab\endcsname\relax\def\natexlab#1{#1}\fi

\bibitem[{Boley(2009)}]{boley09}
Boley, A.~C. 2009, Astrophys. J. Lett., 695, L53

\bibitem[{Boley \& Durisen(2008)}]{boleydur08}
Boley, A.~C. \& Durisen, R.~H. 2008, Astrophys. J., 685, 1193

\bibitem[{Boley \& Durisen(2010)}]{boleydur10}
---. 2010, Astrophys. J., 724

\bibitem[{Boley {et~al.}(2007)Boley, Hartquist, Durisen, \&
  Michael}]{boleyetal07}
Boley, A.~C., Hartquist, T.~W., Durisen, R.~H., \& Michael, S. 2007, Astrophys.
  J. Lett., 656, L89

\bibitem[{Boley {et~al.}(2010)Boley, Hayfield, Mayer, \& Durisen}]{boleyetal10}
Boley, A.~C., Hayfield, T., Mayer, L., \& Durisen, R.~H. 2010, Icarus, 207, 509

\bibitem[{Boss(2003)}]{boss03}
Boss, A.~P. 2003, Astrophys. J., 599, 577

\bibitem[{Cossins {et~al.}(2010)Cossins, Lodato, \& Clarke}]{cossins10}
Cossins, P., Lodato, G., \& Clarke, C. 2010, MNRAS, 401, 2587

\bibitem[{D'Alessio {et~al.}(2001)D'Alessio, Calvet, \& Hartmann}]{d'alessio01}
D'Alessio, P., Calvet, N., \& Hartmann, L. 2001, Astrophys. J., 553, 321

\bibitem[{Durisen {et~al.}(2007)Durisen, Boss, Mayer, Nelson, Quinn, \&
  Rice}]{durisetal07}
Durisen, R.~H., Boss, A.~P., Mayer, L., Nelson, A.~F., Quinn, T., \& Rice, W.
  K.~M. 2007, in Protostars and Planets V, ed. .~K.~K. B.~Reipurth, D.~Jewitt,
  607--622

\bibitem[{Haghighipour \& Boss(2003{\natexlab{a}})}]{hagboss03b}
Haghighipour, N. \& Boss, A.~P. 2003{\natexlab{a}}, Astrophys. J., 598, 1301

\bibitem[{Haghighipour \& Boss(2003{\natexlab{b}})}]{hagboss03a}
---. 2003{\natexlab{b}}, Astrophys. J., 583, 996

\bibitem[{Hayfield {et~al.}(2010)Hayfield, Mayer, Wadsley, \&
  Boley}]{hayfield10}
Hayfield, T., Mayer, L., Wadsley, J., \& Boley, A.~C. 2010, MNRAS, submitted
  astro-ph.EP 1003.2594

\bibitem[{Johnson \& Gammie(2003)}]{jongam03}
Johnson, B.~M. \& Gammie, C.~F. 2003, Astrophys. J., 597, 131

\bibitem[{Mayer(2010)}]{mayer10}
Mayer, L. 2010, in Formation and {Evolution of Exoplanets}, ed. R.~Barnes
  (Wiley-VCH Verlag GmbH \& Co. KGaA)

\bibitem[{{Mayer} {et~al.}(2007){Mayer}, {Lufkin}, {Quinn}, \&
  {Wadsley}}]{mayer07}
{Mayer}, L., {Lufkin}, G., {Quinn}, T., \& {Wadsley}, J. 2007, Astrophys. J.
  Lett., 661, L77

\bibitem[{Mayer {et~al.}(2004)Mayer, Quinn, Wadsley, \& Stadel}]{mayer04}
Mayer, L., Quinn, T., Wadsley, J., \& Stadel, J. 2004, Astrophys. J., 609, 1045

\bibitem[{Mekler \& Podolak(1994)}]{mekler94}
Mekler, Y. \& Podolak, M. 1994, Planet. \& Sp. Sci., 42, 865

\bibitem[{Meru \& Bate(2010)}]{merubate10}
Meru, F. \& Bate, M.~R. 2010, MNRAS, 406, 2279

\bibitem[{Podolak \& Mekler(1997)}]{podmek97}
Podolak, M. \& Mekler, Y. 1997, Plan. \& Sp. Sci., 45, 1401

\bibitem[{Podolak \& Zucker(2004)}]{podolak04}
Podolak, M. \& Zucker, S. 2004, Meteor. \& Planet. Sci., 39, 1859

\bibitem[{Pollack {et~al.}(1994)Pollack, Hollenbach, Beckwith, Simonelli,
  Roush, \& Fong}]{pollack94}
Pollack, J.~B., Hollenbach, D., Beckwith, S., Simonelli, D.~P., Roush, T., \&
  Fong, W. 1994, Astrophys. J., 421, 615

\bibitem[{Pollack {et~al.}(1985)Pollack, McKay, \& Christofferson}]{pollack85}
Pollack, J.~B., McKay, C.~P., \& Christofferson, B.~M. 1985, Icarus, 64, 471

\bibitem[{Rafikov(2009)}]{rafikov09}
Rafikov, R.~R. 2009, Astrophys. J., 704, 281

\bibitem[{Rice {et~al.}(2005)Rice, Lodato, \& Armitage}]{riceetal05}
Rice, W.~K.~M., Lodato, G., \& Armitage, P.~J. 2005, MNRAS, 364, L56

\bibitem[{Rice {et~al.}(2006)Rice, Lodato, Pringle, Armitage, \&
  Bonnell}]{riceetal06}
Rice, W.~K.~M., Lodato, G., Pringle, J.~E., Armitage, P.~J., \& Bonnell, I.~A.
  2006, MNRAS, 372, L9

\bibitem[{Ruden \& Pollack(1991)}]{rudpol91}
Ruden, S.~P. \& Pollack, J.~B. 1991, Astrophys. J., 375, 740

\bibitem[{Stamatellos \& Whitworth(2009)}]{stamatellos09}
Stamatellos, D. \& Whitworth, A.~P. 2009, MNRAS, 400, 1563

\bibitem[{Toon \& Ackerman(1981)}]{toonack81}
Toon, O.~B. \& Ackerman, T.~P. 1981, J. Appl. Optics, 20, 3657

\bibitem[{van~de Hulst(1957)}]{hulst57}
van~de Hulst, H.~C. 1957, Light Scattering by Small Particles (New York: John
  Wiley \@ Sons)

\bibitem[{Vorobyov \& Basu(2010{\natexlab{a}})}]{vorobyov10b}
Vorobyov, E.~I. \& Basu, S. 2010{\natexlab{a}}, Astrophys. J., 719, 1896

\bibitem[{Vorobyov \& Basu(2010{\natexlab{b}})}]{vorobyov10a}
---. 2010{\natexlab{b}}, Astrophys. J. Lett., 714, L133

\end{thebibliography}

\end{document}